\documentclass[twocolumn]{aastex631-edit}

\usepackage{subfig}

\newcommand{\CfA}{\affiliation{Center for Astrophysics | Harvard \& Smithsonian, 60 Garden St., Cambridge, MA 02138, USA}}
\newcommand{\ISTA}{\affiliation{Institute of Science and Technology Austria,
Am Campus 1, 3400 Klosterneuburg, Austria}}
\newcommand{\Caltech}{\affiliation{Division of Physics, Mathematics and Astronomy, California Institute of Technology, Pasadena, CA91125, USA}}
\newcommand{\CaltechTwo}{\affiliation{Cahill Center for Astronomy and Astrophysics, California Institute of Technology, 1200 E. California Blvd. MC 249-17, Pasadena, CA 91125, USA}}
\newcommand{\Purdue}{\affiliation{Department of Physics and Astronomy, Purdue University, 525 Northwestern Avenue, West Lafayette, IN 47907, USA}}
\newcommand{\Valparaiso}{\affiliation{Departamento de Física,
Universidad Técnica Federico Santa María, Avenida España 1680, Valparaíso, Chile}}
\newcommand{\NPF}{\affiliation{Millennium Nucleus for Planet Formation, NPF, Valparaıso, Av. España 1680, Chile}}
\newcommand{\WMKeckObs}{\affiliation{W. M. Keck Observatory,  Kamuela, Hawaii, USA}}
\newcommand{\Princeton}{\affiliation{Department of Astrophysical Sciences, Princeton University, Peyton Hall, Princeton, NJ 08544, USA}}
\newcommand{\Carnegie}{\affiliation{The Observatories of the Carnegie Institution for Science, 813 Santa Barbara St., Pasadena, CA 91101, USA}}
\newcommand{\TAPIR}{\affiliation{TAPIR, Walter Burke Institute for Theoretical Physics, 350-17, Caltech, Pasadena, CA 91125, USA}}

\usepackage{url}
\usepackage{footnote}

\begin{document}

\title{Expansion properties of the young supernova type Iax remnant Pa 30 revealed}

\author[0000-0001-7296-3533]{Tim Cunningham}\thanks{NASA Hubble Fellow}\thanks{tim.cunningham@cfa.harvard.edu}
\CfA

\author[0000-0002-4770-5388]{Ilaria Caiazzo}
\ISTA
\Caltech

\author[0000-0001-5847-7934]{Nikolaus Z.\ Prusinski}
\CaltechTwo

\author[0000-0002-4544-0750]{James Fuller}
\Caltech

\author[0000-0002-7868-1622]{John C. Raymond}
\CfA

\author[0000-0001-5390-8563]{S. R. Kulkarni}
\Caltech

\author[0000-0002-0466-1119]{James D. Neill}
\Caltech

\author[0000-0001-7626-9629]{Paul Duffell}
\Purdue

\author[0000-0002-8650-1644]{Chris Martin}
\Caltech

\author[0000-0002-2398-719X]{Odette Toloza}
\Valparaiso
\NPF

\author[0000-0002-9003-484X]{David Charbonneau}
\CfA

\author[0000-0003-0214-609X]{Scott J. Kenyon}
\CfA

\author{Zeren Lin}
\Caltech

\author[0000-0003-2821-1750]{Mateusz Matuszewski}
\Caltech

\author[0000-0003-2064-4105]{Rosalie McGurk}
\WMKeckObs

\author[0000-0002-1633-6495]{Abigail Polin}
\Carnegie
\TAPIR
\Purdue

\author[0000-0003-3024-7218]{Philippe Z. Yao}
\Princeton



\begin{abstract}
The recently discovered Pa~30 nebula, the putative type Iax supernova remnant associated with the historical supernova of 1181 AD, shows puzzling characteristics that make it unique among known supernova remnants. In particular, Pa~30 exhibits a complex morphology, with a unique radial and filamentary structure, and it hosts a hot stellar remnant at its center, which displays oxygen-dominated, ultra-fast winds.
Because of the surviving stellar remnant and the lack of hydrogen and helium in its filaments, it has been suggested that Pa~30 is the product of a failed thermonuclear explosion in a near- or super-Chandrasekhar white dwarf, which created a sub-luminous transient, a rare sub-type of the Ia class of supernovae called type Iax.
We here present a detailed study of the 3D structure and velocities of a full radial section of the remnant. The Integral Field Unit (IFU) observations, obtained with the new red channel of the Keck Cosmic Web Imager spectrograph, reveal that the ejecta are consistent with being ballistic, with velocities close to the free-expansion velocity. 
Additionally, we detect a large cavity inside the supernova remnant and a sharp inner edge to the filamentary structure, which coincides with the outer edge of a bright ring detected in infrared images. Finally, we detect a strong asymmetry in the amount of ejecta along the line of sight, which might hint to an asymmetric explosion. Our analysis provides strong confirmation that the explosion originated from SN\,1181.
\end{abstract}


\section{Introduction} \label{sec:intro}
The ``Guest Star'' recorded in 1181\,AD by Chinese and Japanese astronomers is one of only five confirmed Galactic supernovae (SNe) recorded in human history. Until recently, SN 1181 was the youngest Galactic supernova without an associated remnant identified with confidence. In 2013, the nebula Pa~30 was discovered by amateur astronomer Dana Patchick in the Wide-field Infrared Survey Explorer (\textit{WISE}; \citealt{wise2010}) archive \citep{cutri2012,parker2016} and there is mounting evidence, including its age and sky location, that this nebula is the supernova remnant associated with SN 1181 \citep{ritter2021,schaefer2023}. 

\begin{figure}
	\centering    
\includegraphics[width=0.4\textwidth]{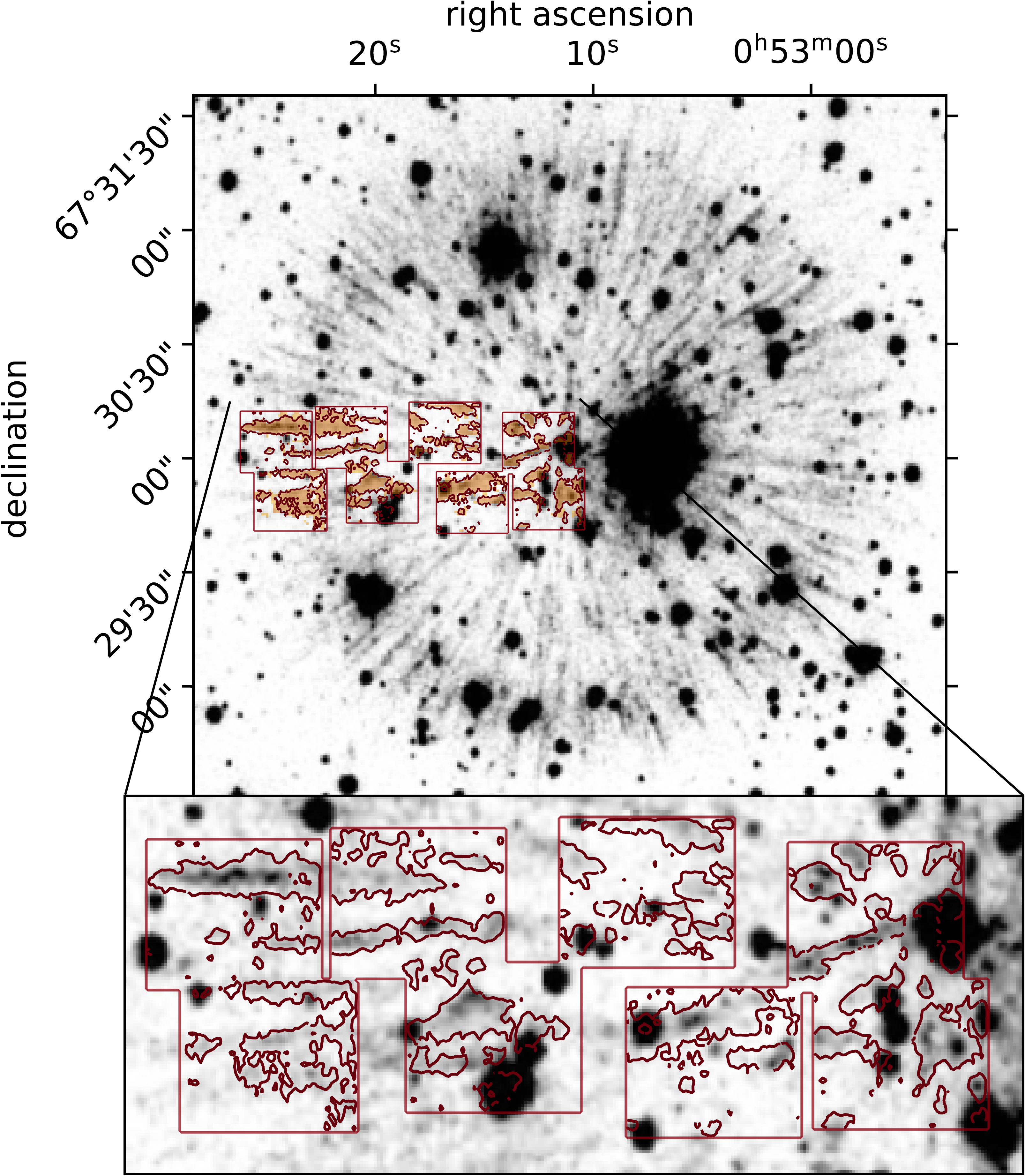}
\includegraphics[width=0.45\textwidth]{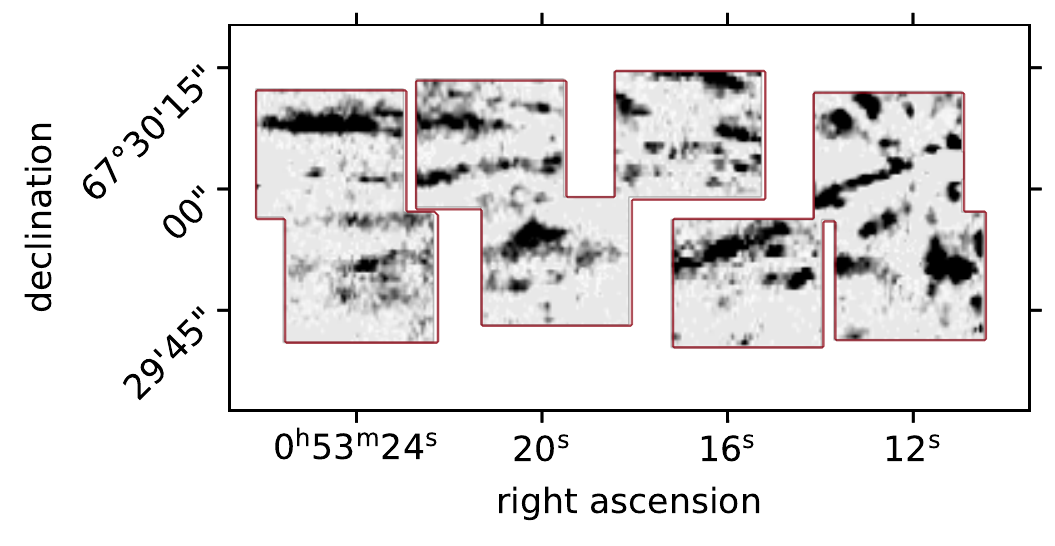}
\caption{ {\it Lower panel}: Narrow-band image obtained from stacking all the KCWI-red channel cubes and integrating over the wavelength range 6680--6750\,\AA, covering the maximally blue- and red-shifted [S\,{\sc ii}] emission features. {\it Upper panel}: the black and white image shows the [S\,{\sc ii}] narrow-band data from \citet{fesen2023}, obtained with the Hiltner telescope at Kitt Peak. Superimposed in orange and with a large transparency, we show for comparison the same KCWI-red image as in the lower panel. The zoomed-in panel shows in red the contours of the filaments in the KCWI-red image: the same filaments are detected in the two images, but the KCWI-red image is deeper, allowing to detect fainter features.}
	\label{fg:Sii-image}
\end{figure}

Pa~30 is a unique SN remnant for several reasons. A hot
star (IRAS 00500+6713, 2MASS J00531123+6730023), with a surface temperature of about 200,000\,K, is 
at its center and shows strong outflows with extreme velocities in excess of 15,000 km/s \citep{gvaramadze2019,lykou2023}. 
Both spectra of the central star and of the nebula lack any indication of hydrogen or helium, while an X-ray spectrum of the nebula reveals carbon-burning ashes \citep{oskinova2020}. Also, the morphology of the SN remnant, revealed by narrow-band images in [S\,{\sc ii}]  \citep[see Fig.\,\ref{fg:Sii-image}]{fesen2023}, has a structure never seen before in a SN remnant, with narrow, almost radial filaments distributed in a spherically symmetric fashion around the central star.

Because of the presence of a surviving stellar remnant and the composition of the star's wind and of the nebula, it has been suggested that Pa~30 is the product of a thermonuclear runaway reaction in a near- or super-Chandrasekhar white dwarf that failed to produce a detonation or failed to explode the entire star, producing a sub-luminous transient \citep{gvaramadze2019,oskinova2020,ritter2021,lykou2023,fesen2023}. This type of transient has been associated to the sub-class of type Ia supernovae called type Iax. Type Iax supernovae also show lower expansion velocities near maximum than normal type Ia, which is consistent with the low velocities detected in Pa~30 \citep{foley2013}. It has been suggested that the Iax supernova was caused by a double-degenerate white dwarf merger \citep{gvaramadze2019,yao2023}, and that the high-speed winds on the surface of the remnant are magnetically driven \citep{gvaramadze2019,zhong2024}.

The morphology of the nebula is strikingly different from the clumpy internal structures typically observed in type Ia remnants, which are caused by a combination of Rayleigh-Taylor instability (RTI) and Kelvin-Helmholtz instability (KHI). On the other hand, examples of somewhat radial filamentary structures have been observed in the wind-blown tails of novae, like GK Persei and DQ Herculis \citep{shara2012,vaytet2007}, and as cometary tails in the infrared images of some planetary nebulae \citep{odell1996,wesson2024}. In analogy to these systems, \citet{fesen2023} suggested that the unusual filamentary structure in Pa~30 is connected to the high velocity wind of the central star, that is accelerating parts of the clumpy, low-density ejecta, and morphing them into filaments by the effect of RTI \citep[see also ][]{ ko2023}. 

However, in the above examples, it is always possible to identify a clump undergoing visible mass loss, while there is no obvious density variation or clumps in Pa~30's filaments. 
Also, the wind-driven scenario does not explain why the turbulence generated by KHI does not destroy the filaments. In an alternative model, \citet{duffell2024} proposed that the filaments are created by RTI in the interaction between the SN ejecta and the circumstellar medium (CSM), and that the KHI in the filaments is suppressed via efficient line cooling. They were able to reproduce the radial structure of the filaments in Pa~30, and made a few predictions on what could be observed in the remnant. In particular, they found the velocity structure in the filaments to be nearly ballistic.

The peculiarity of Pa~30's morphology is also revealed in the infrared and in X-rays. The \textit{WISE} image in the W4 filter centered at 22\,$\micron$ shows a diffuse infrared halo with an angular size of about 100 arcseconds in radius, and a smaller, brighter ring extending to about an arcminute \citep{gvaramadze2019,fesen2023}. \citet{gvaramadze2019} suggested that the infrared nebular emission could be dominated by [O\,{\sc iv}] 25.89\,$\mu$m and [Ne\,{\sc v}] 14.32 and 24.32\,$\mu$m lines. 
On the other hand, photoionization modeling performed by \citet{yao2023} shows that if the progenitor of the Pa~30 remnant is a double C/O white dwarf merger, the measured fluxes in the WISE filters W3 (12\,$\mu$m) and W4 (22\,$\mu$m) are likely to be dominated by dust emission, rather than line emission (see their Figure 8).
\citet{lykou2023} also performed photoionization modeling of the infrared spectral energy distribution (SED) of the nebula out to longer IR wavelengths covering the 12--160\,$\mu$m range, as measured by WISE, IRAS and AKARI (see their Fig.\,12). Their results suggested that the emission at shorter wavelengths ($\lesssim 30$\,$\mu$m) is likely dominated by strong emission lines of [Ne\,{\sc vi}], [Ne\,{\sc v}], and [O\,{\sc iv}] at 7.6\,$\mu$m, 14.7\,$\mu$m, and 25.9\,$\mu$m, respectively, as proposed by \cite{gvaramadze2019}, whilst the SED at longer wavelengths cannot be explained by emission lines but is consistent with emission from dust. JWST spectroscopy and narrow-band imaging would help reveal the origin of the infrared emission in this peculiar source.

In X-rays, the system has been imaged using \textit{XMM-Newton} \citep{jansen2001-xmm} and the \textit{Chandra X-ray Observatory} \citep{weisskopf2000-chandra}. An outer nebula is detected with \textit{XMM} to have an angular size comparable to that of the halo, while a much smaller inner nebula is detected as a point source in \textit{XMM} and it is resolved to be about 1.5 arcseconds in \textit{Chandra} 
\citep{oskinova2020,ko2023}. While the edge of the outer X-ray nebula might correspond to the location of the forward shock, where the ejecta is interacting with the unperturbed CSM, the origin of the inner nebula is more uncertain \citep{duffell2024}. It has been suggested that the inner X-ray nebula corresponds to the location of the wind termination shock, where the high-speed winds from the white dwarf run into the much slower ejecta from the supernova \citep{ko2023}. 

We here report Integral Field Unit (IFU) spectroscopic observations of the Pa~30 nebula that reveal a rich morphology in the SN ejecta and provide the first three-dimensional characterization of the velocity and spatial structure of the nebula. This allows the first accurate determination of the inner edge of the filaments, providing a key constraint for dynamical models of the ejecta. In Section\,\ref{sec:obs} we describe the observations, in Section\,\ref{sec:integrated} we show the sky location of the strongest emission features, in Section\,\ref{sec:velocity} we present the full velocity map, and in Sections \ref{sec:deprojection}\,\&\,\ref{sec:vdist} we present a deprojection of the velocities. In Section\,\ref{sec:summ} we provide a summary of our findings.

\section{Observations} \label{sec:obs}
We observed a radial section of the nebula (see Fig.\,\ref{fg:Sii-image}) using the blue and the
new red arm of the Keck Cosmic Web Imager \citep[KCWI,][]{morrissey2018} Integral Field Unit (IFU) on the Keck II Telescope on Mauna Kea.  We used the low-dispersion gratings, BL and RL, with the medium slicer on the two arms of the IFU. This configuration yields a field of view (FOV) of $16.5''\times20.4''$, a spatial resolution of 0.7'', and a spectral resolution of $R\simeq1800$ on the blue side and of $R\simeq1600$ on the red side. We used a detector binning of $2\times2$ for all our pointings. Fig.\,\ref{fg:Sii-image} shows the location of our mosaic tiles superimposed on the [S\,{\sc ii}] image from \citet{fesen2023}. For each tile, we obtained a single 2000s exposure with KCWI-blue and six separate 300s exposures with KCWI-red. This combination leads to a similar exposure time in both the blue and red channels and mitigates the high influx of cosmic rays on the red side. For the top-right pointing, which contains the central star, we also obtained a shorter 120\,s exposure to avoid saturation. We will present an analysis of our 
spectrum of the central star in an upcoming paper. More details on the data reduction and analysis are given in Appendix~\ref{sec:app}.

\section{Results}
\subsection{Integrated images and spectra} \label{sec:integrated}

Fig.\,\ref{fg:Sii-image} shows the filamentary structure revealed by integrating the stacked KCWI-red data cubes across the wavelength range 6680--6750\,\AA\, which covers the blue- and red-shifted emission from the [S\,{\sc ii}] doublet, compared to the narrow-band image from \citet{fesen2023}. The top panel shows that the filaments detected with KCWI-red are coincident with some of the brightest filaments in the narrowband imaging. However, the KCWI-red data allows us to detect fainter features, which leads to a higher filling factor than implied by the narrow-band image from \citet{fesen2023}. In the KCWI-red and KCWI-blue data, we also detect filaments with emission lines attributed to [O\,{\sc iii}] (at 5007\,\AA) and [Ar\,{\sc iii}] (at 7136\,\AA). This letter focuses on the velocity structure of the nebula derived from the strongest emission features, the [S\,{\sc ii}] doublet. In a follow-up paper, we will present a detailed analysis of the full dataset, including the other emission lines.

\begin{figure}
	\centering
    \subfloat{\includegraphics[width=0.45\textwidth]{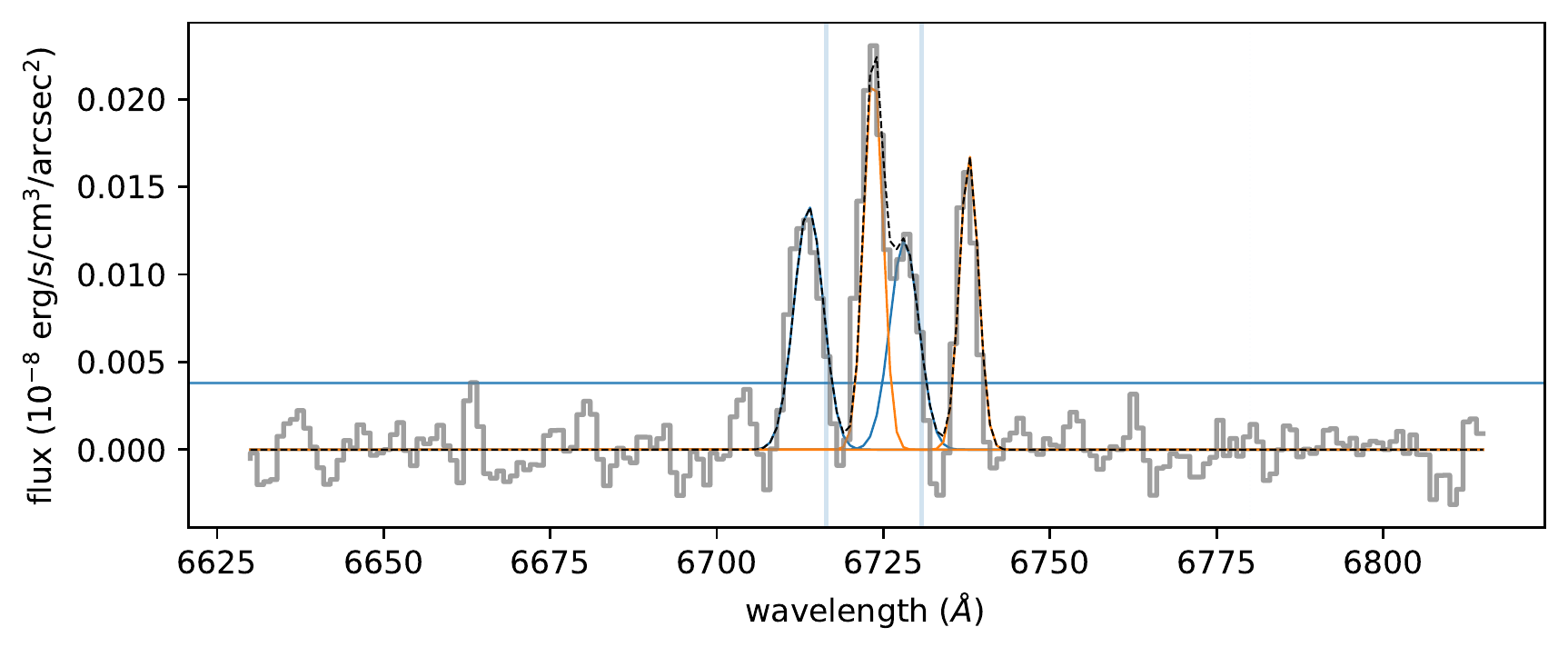}} \\
    \subfloat{\includegraphics[width=0.45\textwidth]{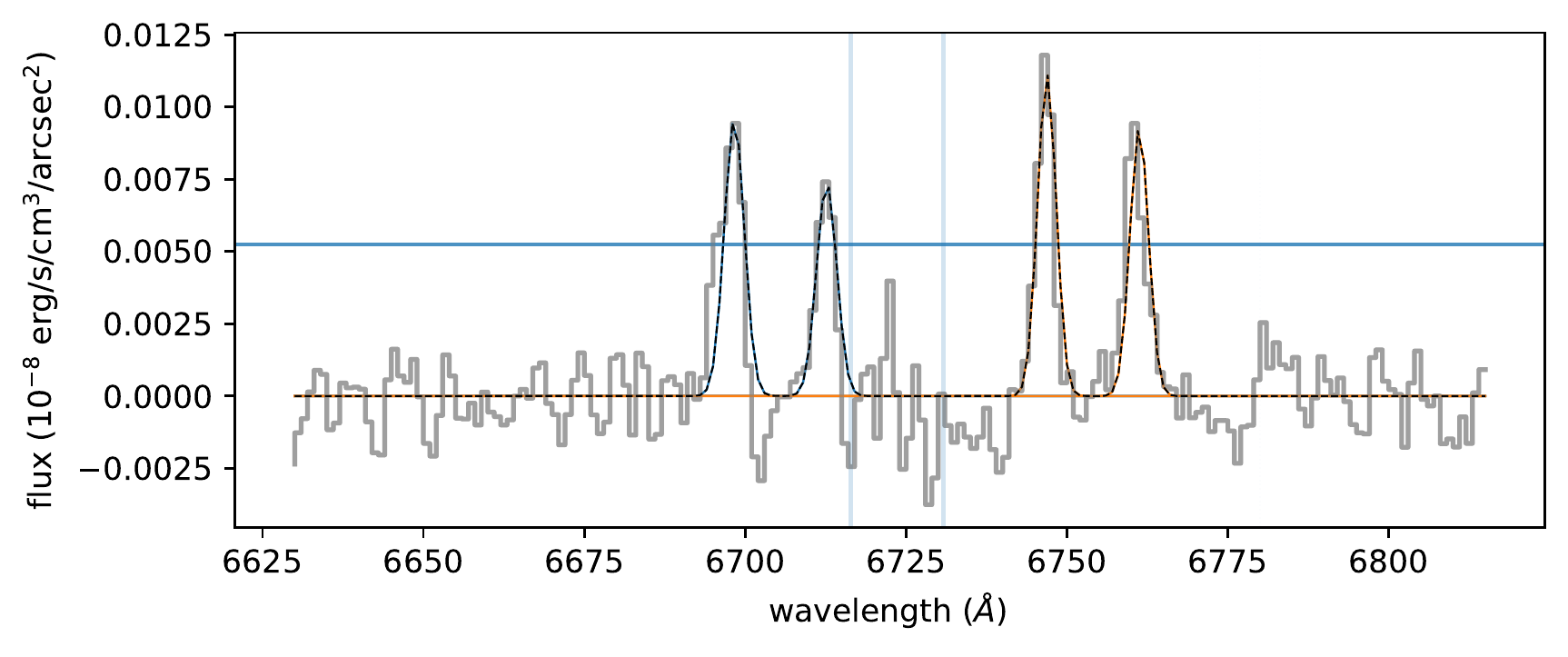}} \\ 
    \caption{Examples of the Gaussian-fitting to the blue- and red-shifted [S\,{\sc ii}] doublets. Each panel shows the mean-averaged spectrum over a 3$\times$3 sky pixel region. The horizontal blue line indicates a 3$\sigma$ noise threshold relative to the local continuum. The two vertical lines indicate the rest wavelengths of the [S\,{\sc ii}] 6717\AA/6731\AA\ doublet.}
	\label{fg:doublet-fits}
\end{figure}

\begin{figure}
	\centering 
\includegraphics[width=0.98\columnwidth]{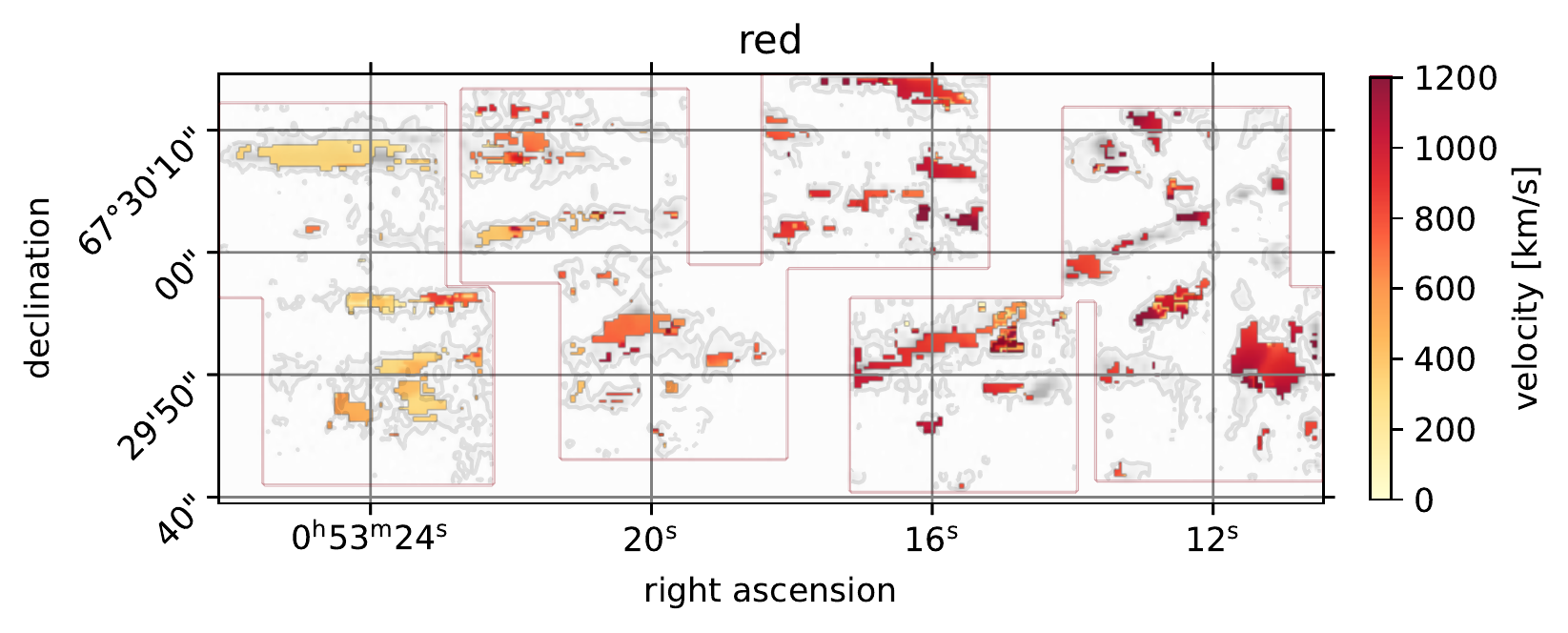}\\
\includegraphics[width=0.98\columnwidth]{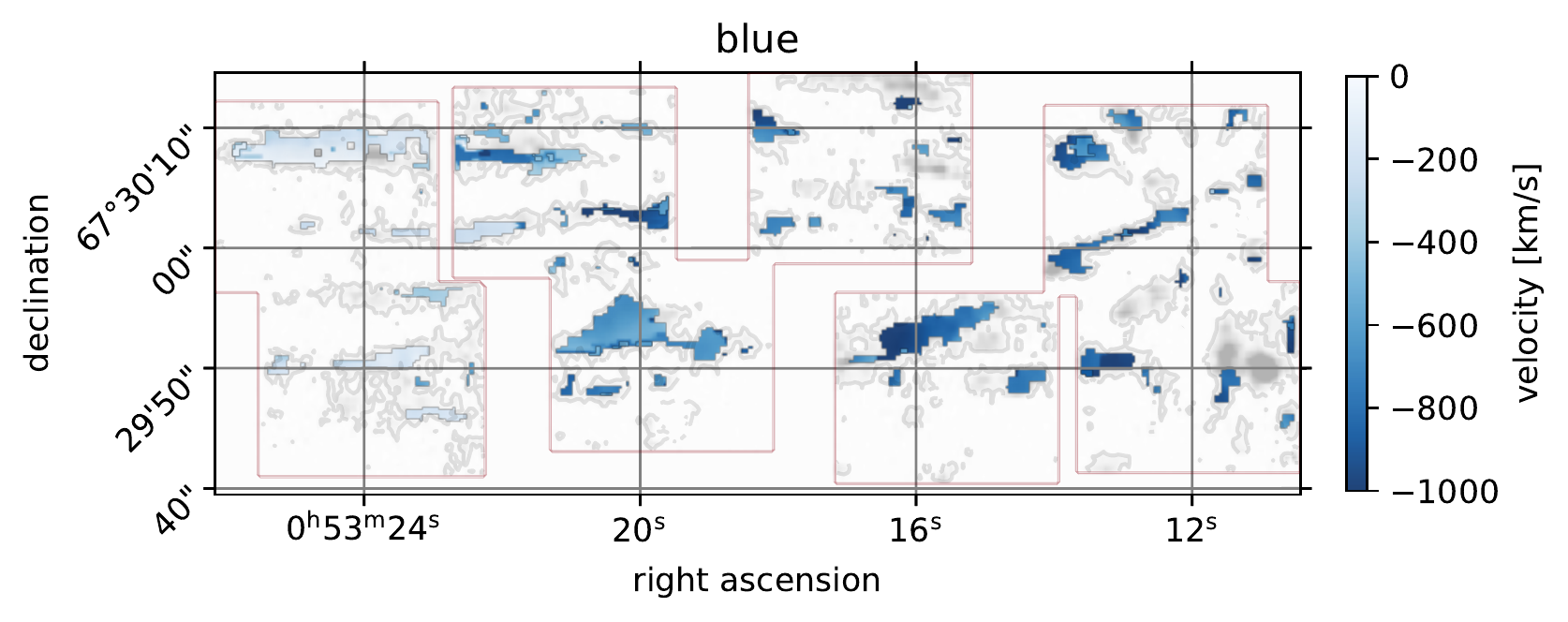}\\
\includegraphics[width=0.98\columnwidth]{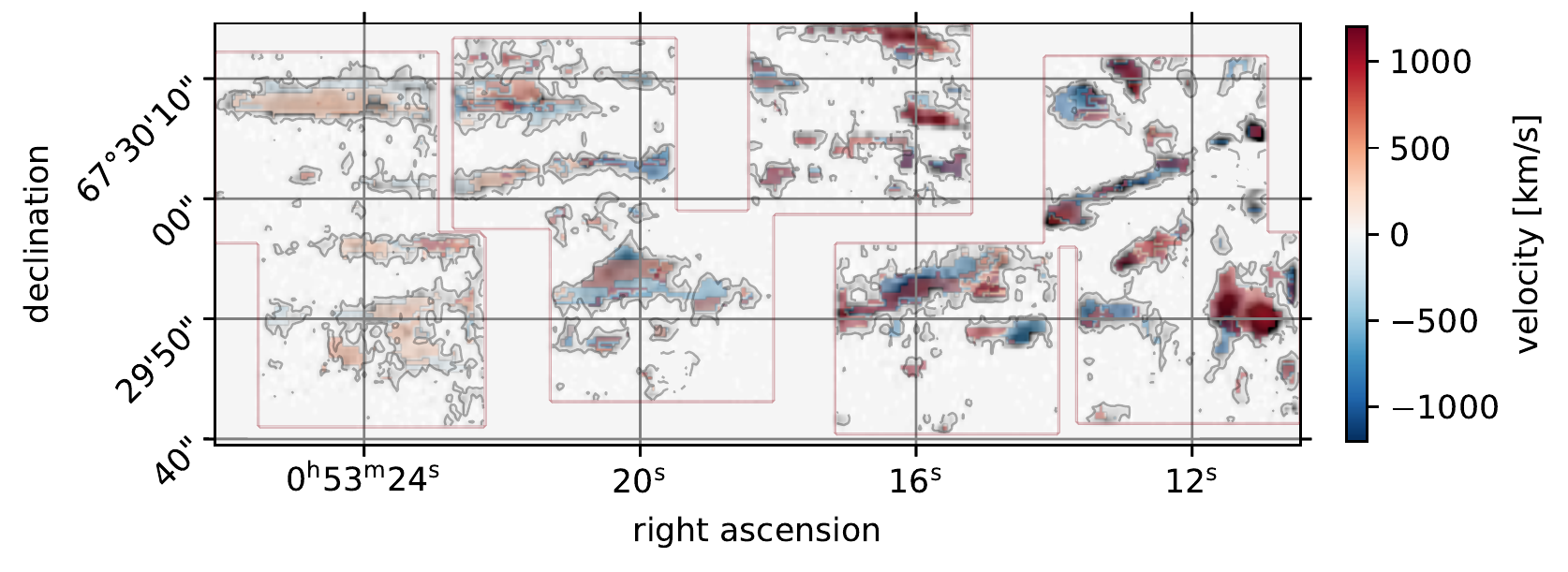}
 \caption{\textit{Upper panel}: velocity map for all sky positions with detectable red-shifted emission in the [S\,{\sc ii}] 6717/6731\,\AA\ doublet. We find a line-of-sight velocity gradient as a function of projected sky distance to the central star, which strongly implies a homologous velocity profile. \textit{Middle}: Similar to the top panel, but for all measured blue-shifted components. \textit{Lower} Combined velocity map for all filaments.}
\label{fg:velocity-map}
\end{figure}

\subsection{Velocity mapping} \label{sec:velocity}
To construct a velocity map of the nebula, we analyze the Doppler shift of the [S\,{\sc ii}] doublet in all filaments.
In order to reduce contamination from noise, we focus only on the spatial regions which lie within the brightest filaments, i.e., within the contours shown in red in Fig.\,\ref{fg:Sii-image}. Details on the production of the mask are given in Appendix~\ref{sec:app}.

At every spatial pixel included within the selected regions, we identify the [S\,{\sc ii}] doublet and perform a double-Gaussian fit to the spectrum in the range 6630--6820\,\AA, mean-averaged over a 3$\times$3 pixel window centered on the given pixel location.
In some pixels, we can find contributions from more than one filament at different velocity shifts. For this reason, we fit additive pairs of Gaussians, either one, two, three or four pairs, to all detected doublets with emission that peaks at least 3$\sigma$ above the local continuum noise level. The two Gaussians in each pair have a fixed wavelength separation of $\Delta\,\lambda=6730.78-6716.42=14.6$\,\AA, and the amplitude ratio is limited to the range $0.5\leq A_{6717}/A_{6731}\leq 1.3$. 
The low- and high-density limits of 1.3 and 0.5 are determined
purely by the atomic rates \citep[see, for instance Figure 9 of ][ for the ratio from the atomic database for emission lines CHIANTI, \citep{delzanna2015}]{seok2020}.
Although the true low density limit is $\approx$1.42, densities much below 100\,cm$^{-3}$ would not produce significant observable [S\,\textsc{ii}] emission, so we use a more restricted range. 
Fig.\,\ref{fg:doublet-fits} shows an example of the spectral fitting. 
The measured wavelength of the doublet provides the Doppler shifted line-of-sight velocity of the filament(s). The spatial dependence of the measured velocities is shown in Fig.\,\ref{fg:velocity-map}. 

Near the central star, we measure absolute velocities up to $\approx1,400$\,km\,s$^{-1}$, higher than the $\approx1,100$\,km\,s$^{-1}$ reported by previous single-slit spectroscopic observations \citep{ritter2021,fesen2023}. We also find strong variation in velocity along the filaments, many of which show a clear increase in velocity as a function of radius. Although some filaments are predominantly red-shifted or blue-shifted, many of the bright filaments appear to show contributions from both blue- and red-shifted components. Such apparent spatial correlation is likely not significant, as we only observed a small fraction of the nebula. To investigate if there is an actual spatial correlation between blue- and red-shifted emission, we performed two statistical tests.

In our fitting procedure we focused on the pixels with the brightest emission. Such selection resulted in fits of 8903 unique [S\,{\sc ii}] doublet. Among those were a total of 4888 red-shifted and 4015 blue-shifted doublets. In total, 5575 unique pixels yielded a fit of at least one doublet, whilst 2894 pixels yielded a fit of two or more doublets. In short, we find that the fraction of pixels with more than one distinct velocity contribution is $2894 / 5575=0.52$, which is not negligible but shows that any given pixel is just as likely to have one rather than multiple velocity contributions. To further test the significance of this apparent correlation we perform a 2D Kolmogorov–Smirnov test on the spatial position of two samples; on sample containing the sky coordinates of all red-shifted fitted doublets, the other sample containing the coordinates of all blue-shifted components. The 2D KS test is performed using the public code \textsc{ndtest}\footnote{Written by Zhaozhou Li, \url{https://github.com/syrte/ndtest}}
which provides a \textsc{python} implementation of the two-dimensional Kolmogorov-Smirnov test algorithm developed by \cite{peacock1983-2dks} and later improved by \cite{fasano1987-2dks}. The $p$-value from the 2D KS test provides an estimate of the probability with which the null hypothesis can be ruled out. In this case, the null hypothesis that the two samples are drawn from the same spatial distribution has a probability of $p<10^{-17}$. We thus conclude that there is likely no significant correlation between the sky position of red- and blue-shifted material in these KCWI observations.

\begin{figure}
	\centering
    \includegraphics[width=0.47\textwidth]{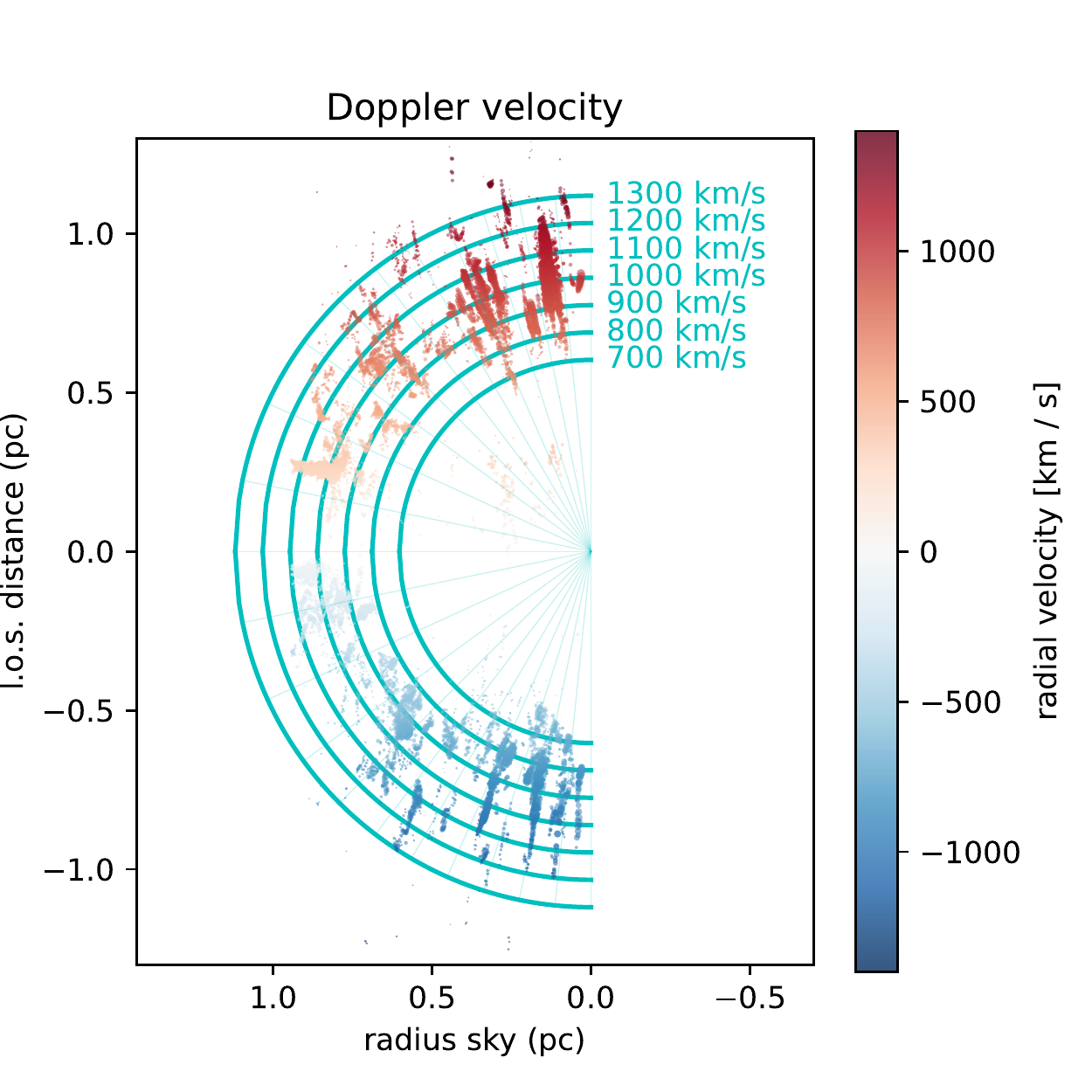}
    \caption{Deprojected radial structure of the observed fraction of the nebula assuming that all material was ejected in the year 1181 and that the velocities have remained constant in time. The $x$-axis shows the radial distance between the central star and a given filament projected on the plane of the sky assuming the \textit{Gaia} distance to the central star of $D$\,=\,$2.3$\,kpc. The $y$-axis shows the re-constructed line-of-sight distance from the central star, using Equation\,\ref{eq:dlos} and an age of {$\tau$\,=\,$842$\,yr}.}
	\label{fg:deproject}
\end{figure}

\begin{figure*}
	\centering
    \subfloat{\includegraphics[width=0.9\textwidth]{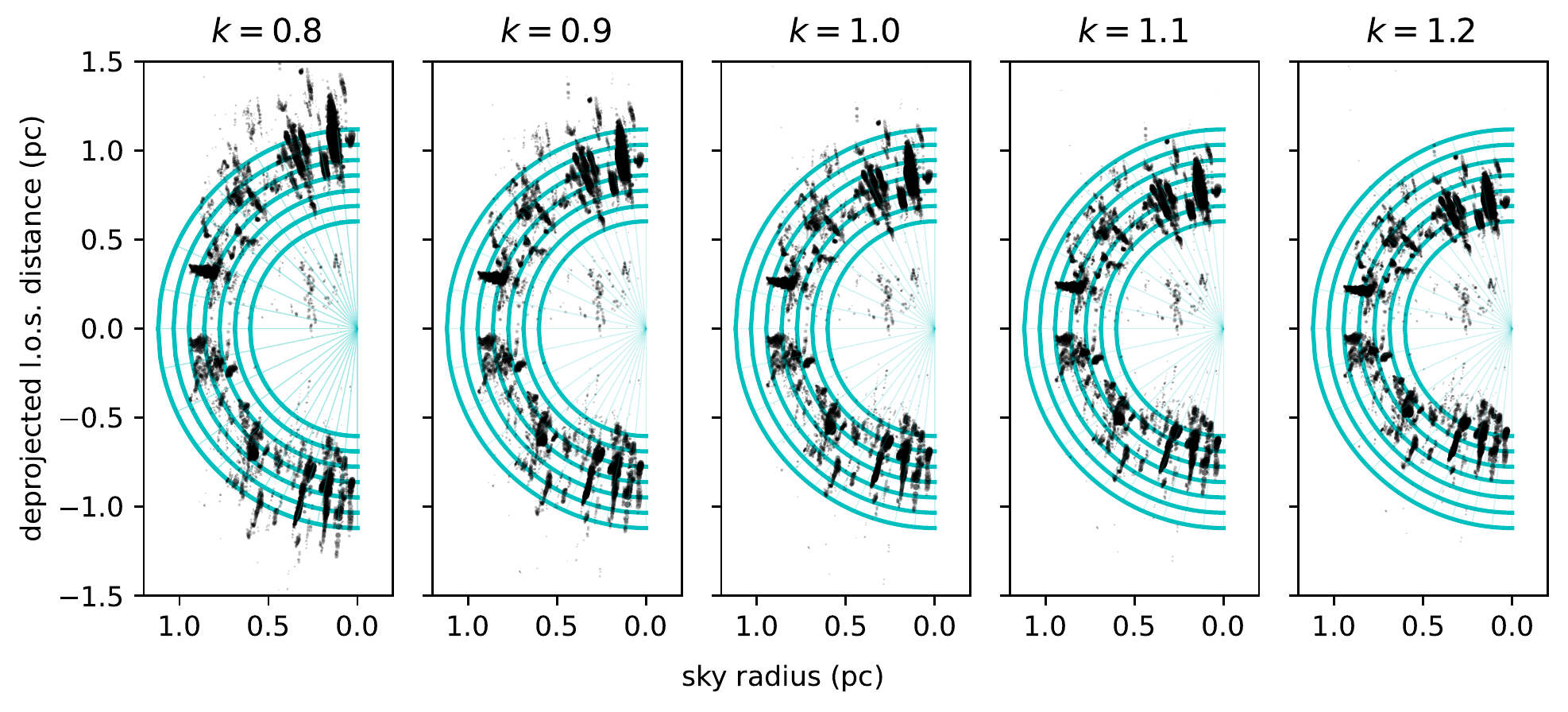}}
	\caption{Deprojection of the measured line-of-sight velocities and sky position. The $x$-axis parameter, sky radius, describes the distance between the central star and any given point on the sky, collapsing the two dimensional sky positions into one dimension only. The parameter $k$, increasing from left to right, indicates the fraction of the ballistic velocity used in the deprojection. Under the assumptions that all material was ejected in the year 1181 and that the nebula has approximately spherical symmetry (as implied by the existing narrowband imaging \cite{fesen2023}), we can see qualitatively that $k$ should be close to unity, and therefore that the velocities in the ejecta are close to the ballistic free-expansion velocity. The contours show the predicted spatial location for material ejected at different velocities, from 700\,km/s at the inner edge, to 1300\,km/s at the outer edge.}
	\label{fg:ballistic}
\end{figure*}

\begin{figure*}
	\centering   
\subfloat{\includegraphics[width=0.48\textwidth]{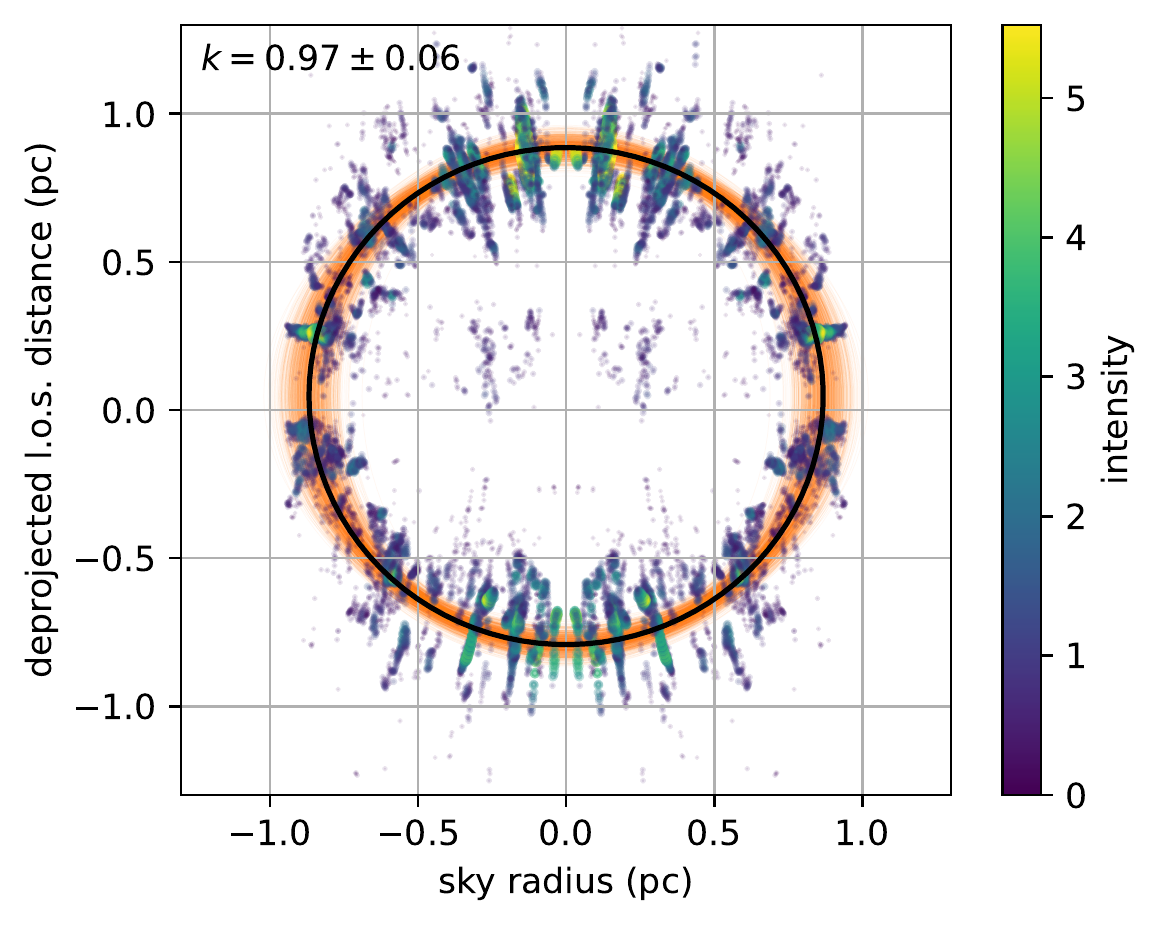}}
\includegraphics[width=0.4\textwidth]{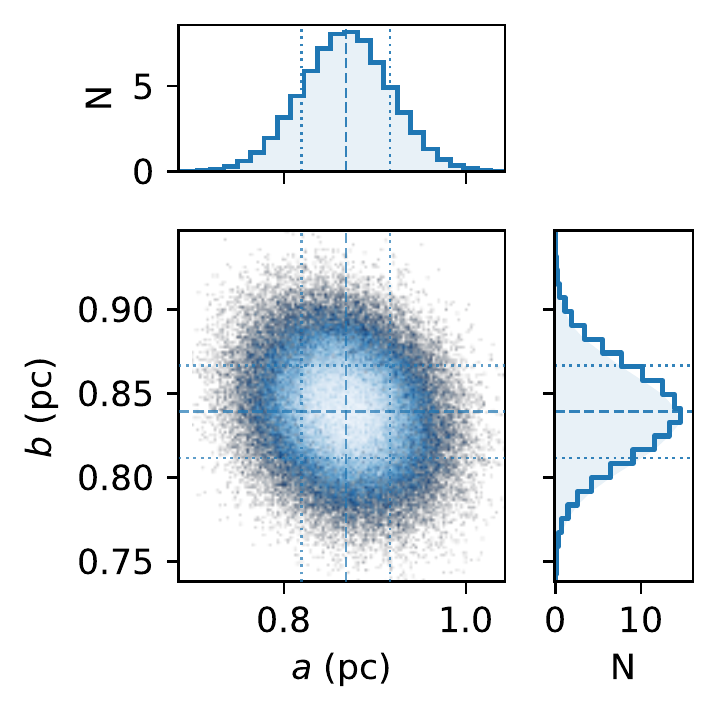}
	\caption{Deprojection fit with the assumptions outlined in the caption of Fig.\,\ref{fg:ballistic}. The measured Doppler shift is converted into a line-of-sight distance using Equation\,\ref{eq:dlos}, and plotted against the measured \textit{sky radius;} which is defined as the distance from the central star on the sky in parsec. The central star is located at the origin of the graph. The data is reflected across the plane defined by $r_{\rm sky}=0$ to enable an elliptical regression. 
    The best-fit deprojection contour (solid) is found using the linear least-squares Levenberg-Marquardt algorithm to fit for the five parameters of the ellipse: $a$, $b$, $x_0$, $y_0$, and $\theta$, which represent the semi-major axis, semi-minor axis, x-y origin and angle. The ballistic fraction, $k$, is then given by $k=b/a$. This yields a best-fit value of $k=0.97^{+0.09}_{-0.08}$, where the 1$\sigma$ uncertainties also account for the uncertainty on the \textit{Gaia} distance. The fitted values for the center of the ellipse are $x_0=0.00\pm0.05$\,pc and $y_0=0.05\pm0.03$\,pc. \textit{Right:} Posterior distribution of $a$ and $b$ from the elliptical regression shown in the left panel.}
	\label{fg:ballistic-fit}
        \label{fg:k-t0-hist}
\end{figure*}

\subsection{Deprojection} \label{sec:deprojection}
The measurement of the sky position and line-of-sight velocity for each filament allows, for the first time, a fully three-dimensional reconstruction of the Pa~30 nebula. 
We adopt the distance to the central star to be $D$\,=\,$2.3^{+0.1}_{-0.1}$\,kpc, as presented in the geometric \textit{Gaia} distance catalog of \cite{gaia2021}.
First, we deproject the structure of Pa~30 by assuming that the measured ejecta velocities have been constant in time since the year 1181, and that all gas with detected emission was ejected simultaneously. Therefore, for each measured line-of-sight velocity, $v_{\rm{los}}$, we find the line-of-sight distance from the central star as 
\begin{equation}
\label{eq:dlos}
    d_{\rm{los}}= v_{\rm{los}}\tau\,~~~,
\end{equation}
where we take the age of the remnant to be $\tau=842$\,yr. In Fig.\,\ref{fg:deproject}, we plot the recovered $d_{\rm{los}}$ as well as the line-of-sight velocities. We can see that the structure in Fig.\,\ref{fg:deproject} is very close to being spherically symmetric. After the deprojection, we can clearly see a cavity near the central star. There is still a small percentage of pixels (2.6\%) redshifted side close to the central star (in the region defined by $r_{\rm sky}<0.5$\,pc and $0.0<d_{\rm los}<0.5$\,pc); visual inspection of the spectra of the 229 pixels that produce doublets in that region reveals that they are mostly due to contamination from a stationary sky line, which was not properly removed for this small percentage of pixels. Because of their small fluxes, these noisy pixels do not affect our analysis.

Since the line-of-sight velocities measured by Doppler shifts indicate the current velocities in the ejecta, while the sky location of the ejecta divided by the time since the explosion carries information on how the velocities have changed since the explosion, it is not obvious that the spatial structure should appear symmetric after the deprojection. In fact, asymmetry between the x and the y axis in Fig.\,\ref{fg:deproject} can be expected in three cases: 1) the assumed age for the remnant is incorrect, 2) the velocities in the nebula are not ballistic, and the ejecta has been slowed down or accelerated since the supernova, or 3) the velocity of the ejecta along the line of sight is intrinsically different to velocities in the sky plane. 

We parameterize the amount by which the filaments deviate from fully ballistic by defining the \textit{ballistic fraction}, $k$, such that
\begin{equation}
\label{eq:v-ballistic}
    v=\frac{kr}{\tau}\,~~~,
\end{equation}
where $r$ is the radial distance from the central star and $\tau$ is the time since the supernova explosion. From the data, we can only constrain the ratio $k/\tau$, as the two parameters are fully degenerate.

Lifting the assumption that the ejecta is homologous, the line-of-sight distance is now given by \begin{equation}
\label{eq:dlos-k}
    d_{\rm{los}}= \frac{v_{\rm{los}}\tau}{k}\,~~~.
\end{equation}
In Fig.\,\ref{fg:ballistic}, we plot the line-of-sight distance converted from the measured Doppler shift by using Equation\,\ref{eq:dlos-k} against the measured \textit{sky radius}, which is defined as the distance from the central star on the sky in parsec. In the different panels, we keep $\tau=842$\,yr and increase, from left-to-right, the ballistic fraction $k$ from 0.8 to 1.2, in order to demonstrate the effects of varying the $k/\tau$ ratio. 
We emphasize that in reality, we are varying the $k/\tau$ ratio as the two parameters are degenerate. 
Qualitatively, we can see that the ejecta look approximately spherically symmetric when we assume $k\sim1.0$.

To quantitatively constrain the ratio $k/\tau$, in Fig.\,\ref{fg:ballistic-fit} we perform an elliptical fit to the spatial distribution of the filaments, reconstructed using Equation\,\ref{eq:dlos}. The datapoints are reflected across the plane defined by $r_{\rm sky}=0$ to enable an elliptical regression. Since the sky radius is by definition positive, this is equivalent to fitting the non-reflected data with a semi-ellipse. However, the full ellipse is computationally convenient and we don't expect this reflection to alter the statistical conclusions from this analysis. 
The best-fit deprojection contour (black solid line) is found using a linear least-squares Levenberg-Marquardt algorithm implemented in the \textsc{scipy} package \textsc{least\_squares}. The fit is performed over five parameters which describe the ellipse: $a$, $b$, $x_0$, $y_0$, and $\theta$, which represent the semi-major axis, semi-minor axis, the x-y origin and the angle of rotation between the semi-major axis and the $x=0$ plane. The fit is performed by minimizing the geometric distance between the datapoints and the ellipse. The geometric distances are weighted by the integrated intensity, $I$, in the two emission lines of the detected [S\,{\sc ii}] doublet (see Fig.\,\ref{fg:doublet-fits}). The integral is described analytically by $I=\sqrt{2\pi}\left((\sigma A)_{\rm 6717} + (\sigma A)_{\rm 6730}\right)$, where $A$ and $\sigma$ are the fitted Gaussian amplitude and standard deviation, respectively. 
The uncertainty on the fit is shown in the figure by sampling the multivariate normal distribution described by the covariance matrix from the least-squares regression. The thousand drawn samples are plotted in orange. 

From the best-fit ellipse parameters, the ratio of the semi-minor and -major axes quantifies the asymmetry in the distribution, with $b/a=1$ indicating a perfect circle. If we assume that our estimate of the age is correct, the ratio of the semi-minor and -major axes can then be directly converted into an estimate of $k$. We find:
\begin{equation}
    k=\frac{b}{a}=0.97^{+0.09}_{-0.08} \times \left(\frac{\tau}{842\, \mathrm{yr}}\right) \left(\frac{D}{2297\, \mathrm{pc}}\right)^{-1}~~~,
\end{equation}
where the quoted 1$\sigma$ uncertainty includes the 1$\sigma$ uncertainty on the fitted parameters and on the \textit{Gaia} distance posterior \cite{gaia2021}. 
On the other hand, assuming that the ejecta has not been slowed down at all by the CSM ($k\equiv1$), nor accelerated by any mechanism, we can derive an age of the remnant, and therefore the time of explosion to be:
\begin{equation}
    t_0=2023-\tau=1152^{+77}_{-75}\,\mathrm{yr}~~~.
\end{equation}
This result is consistent with the nebula being ejected in the year 1181, and provides a strong indication that the nebula is the remnant of the historic supernova.

Finding $k$ close to unity means that the ejecta have not been accelerated or decelerated significantly since the explosion. It is expected that the ejecta should start decelerating once the forward shock has swept up a mass comparable to the ejecta mass itself, and then the deceleration should continue as more mass is swept up until the expansion enters the adiabatic or ``Taylor-Sedov'' phase.
\cite{oskinova2020} derived an ejecta mass of $M_{\rm ej}\sim0.1\,M_{\odot}$ for Pa~30, based on the measured angular size and emission measure of the outer X-ray nebula, which is consistent with the expected ejecta mass from double white dwarf mergers \citep{schwab2021}. 
Given a radius of $\approx$1\,pc, the ejecta would have swept up $\sim$0.1--1\,$M_{\odot}$ given a typical interstellar medium (ISM) particle density of $\sim$1--10\,cm$^{-3}$; however, the planetary nebula(e) that gave rise to the white dwarf(s) might have created a lower-density region around the system.

\begin{figure}
	\centering
\includegraphics[width=0.45\textwidth]{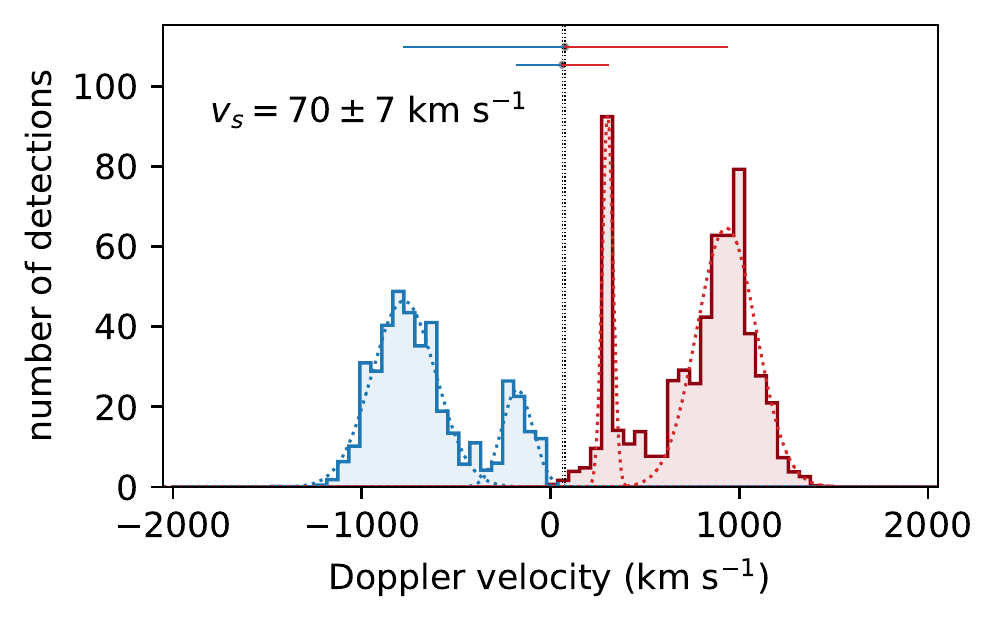} \\
\includegraphics[width=0.45\textwidth]{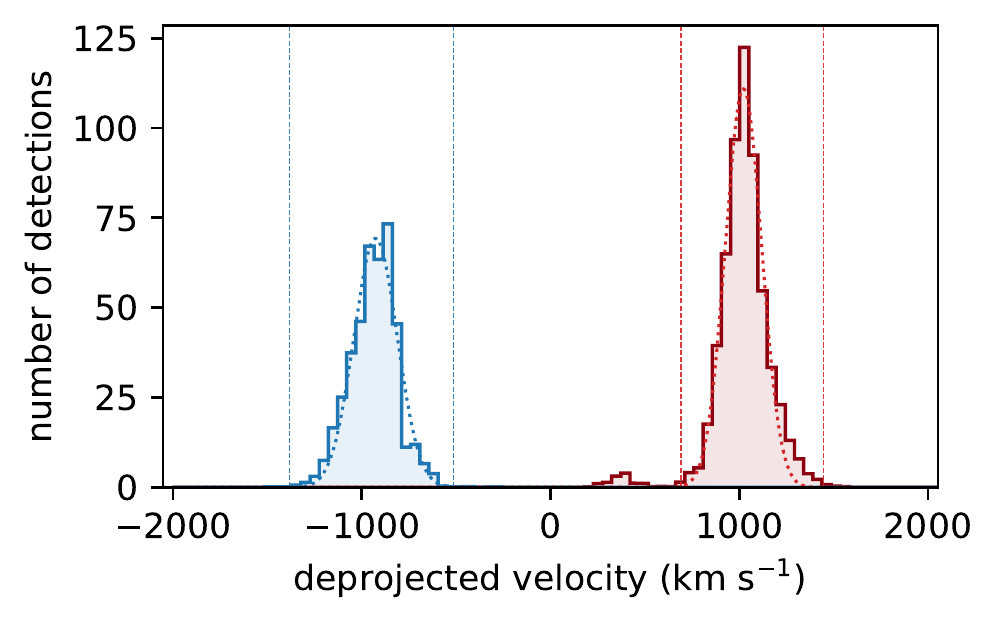}\\
\caption{\textit{Top}: Histogram of the measured Doppler shift for all detected [S\,{\sc ii}] doublets, weighted by the measured flux in each doublet. The four peaks are fit with four Gaussians, finding two high-velocity broad peaks centered on $-$778 and $+$932\,km\,s$^{-1}$, and two narrower, lower-velocity features centered on $-$177 and $+$303\,km\,s$^{-1}$. The peak-to-peak distance between symmetric peaks is shown with the horizontal lines, whilst the vertical dashed line shows the mean velocity between each Gaussian pair. Assuming a spherically-symmetric ejecta, this mean velocity provides an estimate of the total radial velocity of the system, which we find to be $v_{\rm s} = 70\pm7$\,km\,s$^{-1}$. In this figure, all velocities are given in the rest frame of Earth. The heliocentric correction is found to be $v_{\rm H}=+8.08$\,km\,s$^{-1}$, yielding a total system line-of-sight velocity in the rest frame of the Sun to be $v_{\rm sys}=78\pm4$\,km\,s$^{-1}$. \textit{Bottom}: Velocity distribution for the deprojected velocities computed via Equations\,\ref{eq:v-ballistic}\,\&\,\ref{eq:v-pythagoras}. The distribution becomes almost fully bimodal, with peaks at $\pm1000$\,km\,s$^{-1}$. The fitted central velocities are $\overline{v_{\rm b}}=-927\pm 4$ and $\overline{v_{\rm r}}=1022\pm 2$\,km\,s$^{-1}$, for the blue and red shifted components, respectively. The vertical dashed lines indicate the 99\% confidence interval on the filament velocities. 
}
	\label{fg:velocity-histogram}
\end{figure}

\subsection{Velocity distribution} \label{sec:vdist}
In the top panel of Fig.\,\ref{fg:velocity-histogram}, we show the distribution of the measured line-of-sight velocities (given in the rest frame of the Earth) for all detected filaments, weighted by the intensity. We find the multi-modal distribution to have two distinct pairs of symmetric peaks. We fit Gaussians to the peaks and find that the two dominant high-velocity peaks are centered around $-778$ and $+932$\,km\,s$^{-1}$, whilst the two narrower, lower-
velocity features have fitted centers at $-177$ and 
$+303$\,km\,s$^{-1}$. Assuming symmetry along the line of sight, this allows the determination of the total radial velocity of the system. The mean of the Doppler velocities of the high-velocity blue- and red-shifted features is 77\,km\,s$^{-1}$, while for the low velocity features is 63\,km\,s$^{-1}$. We can therefore estimate the total system radial velocity as $v_{\rm sys}=70\pm7$\,km\,s$^{-1}$. Given the observational epoch (16$^{\rm th}$ October, 2023), we compute the heliocentric correction using \textsc{astropy} to be $v_{\rm H}=8.08$\,km\,s$^{-1}$. We thus find a total system line-of-sight velocity in the rest frame of the Sun to be $v_{\rm sys}=78\pm4$\,km\,s$^{-1}$.

From Eq.~\ref{eq:dlos-k}, we can infer a deprojected, ``true'', three-dimensional separation of each filament from the central star $d_{\rm pro}$, given by
\begin{equation}
\label{eq:d-pythagoras}
    d_{\rm pro} = \sqrt{d_{\rm los}^2 + {r_{\rm sky}^2}}~~~.
\end{equation}
In turn, this yields a deprojected ``true'' velocity as
\begin{equation}
\label{eq:v-pythagoras}
    v_{\rm pro} = \frac{d_{\rm pro}}{\tau}~~~.
\end{equation}
In the lower panel of Fig.\,\ref{fg:velocity-histogram}, we show the distribution of all deprojected velocities. We find that the low velocity features visible in the top panel disappear almost entirely as the deprojection pushes them to higher ``true'' velocities, because those filaments lay further away from the star, near the edge of the nebula. The red-shifted low-velocity bump that remains corresponds to the small percentage (2.6\%) of detected doublets that are due to contamination from a stationary sky line (see section\,\ref{sec:deprojection}). We find that, given the small number of doublets in that region, and their generally small fluxes, these features have a negligible impact on the elliptical fit. 
We thus find strong evidence that the nebula is expanding with a mean velocity of $\approx$1000\,km\,s$^{-1}$. The standard deviation of the Gaussians is 115 and 96\,km\,s$^{-1}$, for the blue and red components, respectively. As a measure of the range of velocities in the filaments, the vertical dashed lines in the bottom panel of Fig.\,\ref{fg:velocity-histogram}, indicate the 
99\% confidence interval on the filament velocities. We find the maximum red- and blue-shifted velocities to be 1,440 and 1,380\,km\,s$^{-1}$ respectively, whilst the minima are 650 and 
510\,km\,s$^{-1}$.

From both panels of Fig.\,\ref{fg:velocity-histogram} we can clearly see that the size and intensity of red-shifted filaments is higher than for blue-shifted ones. By separately integrating the flux in all red- and blue-shifted  filaments, we find that there is 40\% more flux measured in red-shifted emission in the $\approx$10\% of the nebula covered by our KCWI-red observations. As the red-shifted filaments lie farther away from us, this cannot be due to a selection effect and may hint at evidence of asymmetry in the ejecta.  In their analysis of a single long-slit spectrum placed at a position angle 30$^\circ$ east of north and close to the central star;  \citet{ritter2021} found the opposite to be true, i.e. a higher flux in the blue-shifted emission. Although the area covered by the long-slit spectrum is much smaller than the one covered by our IFU observations, the discrepancy might indicate that there is variation across the nebula on the plane of the sky. A wider coverage of the nebula is needed to confirm the possible asymmetry in the ejecta. 

\begin{figure*}
	\centering
 \includegraphics[width=0.49\textwidth]{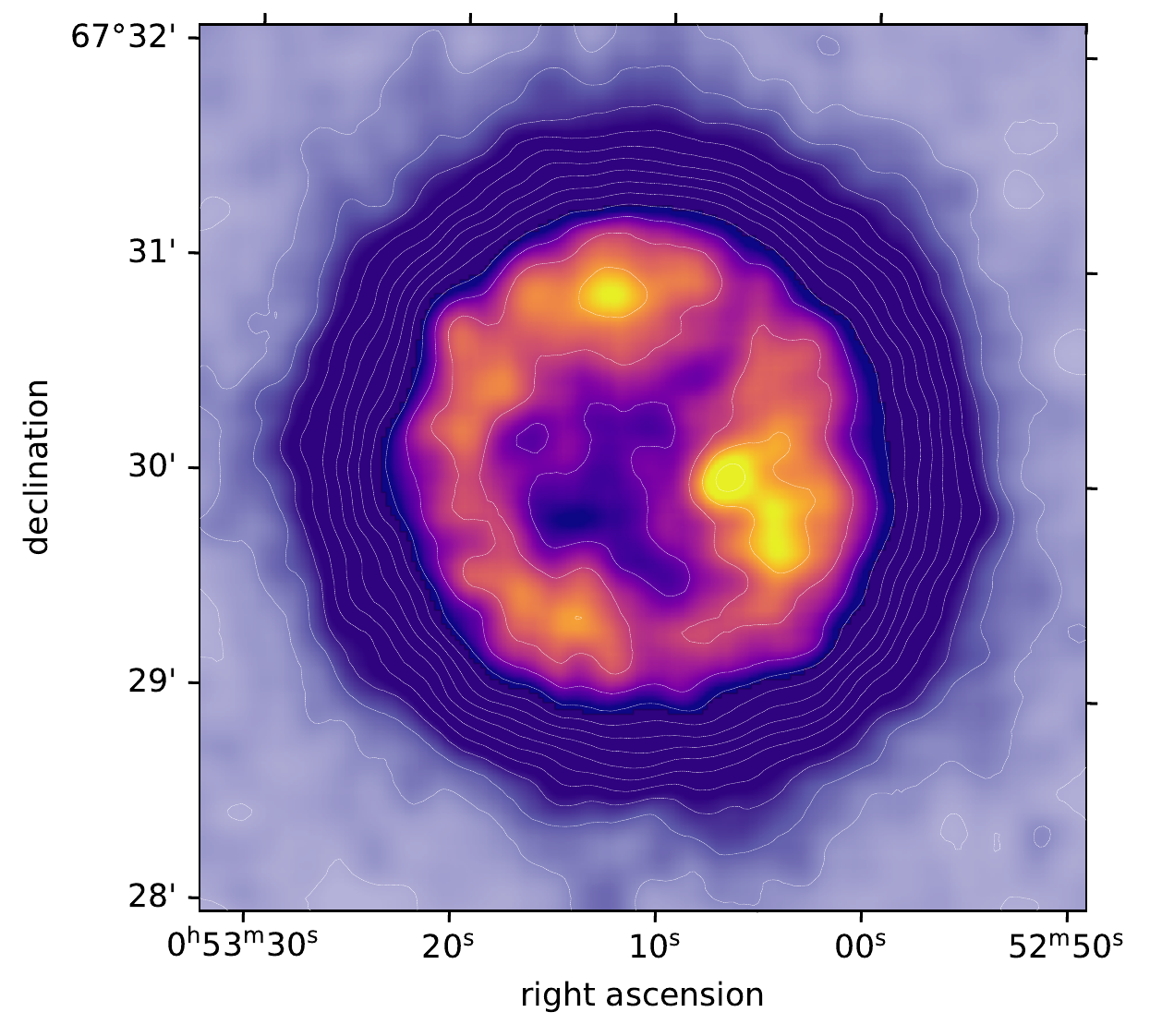}
\includegraphics[width=0.49\textwidth]{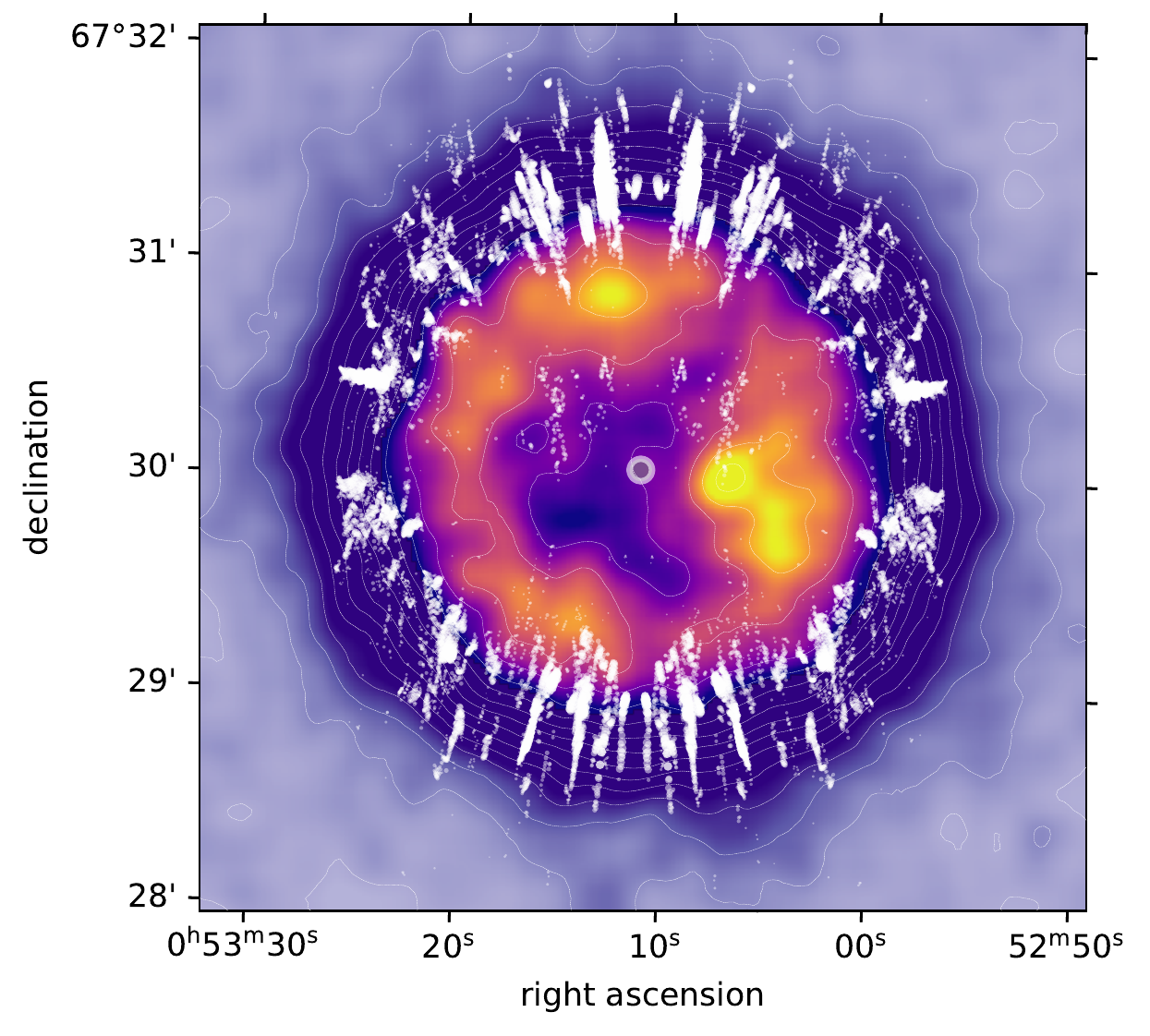} \\
\caption{Left: \textit{WISE} W4 (22\,$\micron$) image from the \textit{WISE} image service \citep{wise2010}. Because the inner infrared nebula in the WISE image is much brighter than the outer one, to show the location of both nebulae we superimpose two versions of the same image with different contrast levels, following the method of \cite{gvaramadze2019}, so that the outer nebula appears in uniform dark purple. For clarity, we also overlay contours of constant W4 infrared flux (thin white lines). The 14 contours plotted are evenly-spaced in flux, with a separation of 8\,$\mu$Jy, from 5.44--5.54\,mJy. The minimum and maximum flux in the image are 5.43 and 5.55\,mJy, respectively.
Right: The WISE image is the same as in the left plot. On top of the WISE image, we plot as white points the deprojected positions of the filaments using our best-fit value of the ballistic fraction, $k$. The deprojected positions have been reflected, as they are in Fig.\,\ref{fg:ballistic-fit}, and rotated by 90 degrees towards the viewer, turning line-of-sight positions into sky positions, whilst preserving the angular size. 
}
	\label{fg:wise-overplot}
\end{figure*}

\subsection{Comparison with infrared imaging}
Based on the imaging from \textit{WISE}, \cite{gvaramadze2019} defined the edge of the bright inner infrared region to be $\approx$1\,arcmin. At the \textit{Gaia} distance to this system, this corresponds to approximately 0.67\,pc in radius. In our dataset, we find that all of the filaments that appear close to the central star in the plane of the sky show high Doppler shifts, implying that their separation from the star is large along the line of sight. We therefore find a cavity in the center of the supernova remnant, which was already noted in \citet{fesen2023}, and a sharp inner edge to the filamentary structure. The location of this inner edge is consistent with the outer edge of the bright, inner infrared ring (see Fig.\,\ref{fg:wise-overplot}).  

In Fig.\,\ref{fg:wise-overplot} we show the \textit{WISE} W4 image at 22\,$\micron$. Following the methodology of \cite{gvaramadze2019}, we plot the \textit{WISE} image at two different intensity scales. This allows us to highlight the edges of the bright inner region, and the less bright outer nebula. In order to aid interpretation of the \textit{WISE} W4 image, contours of constant flux are also shown in the figure using a global/continuous intensity scale. The 14 contours shown are separated in flux by 8\,$\mu$Jy, from the background level of 5.43\,mJy. In order to make a comparison between the spatial position of the filaments detected in this work, and the infrared nebula, we plot the deprojected line-of-sight positions of all measured [S\,{\sc ii}] emission, using our best-fit value of $k=0.97$. The deprojected line-of-sight positions are converted into an effective declination, which is equivalent to a rotation of 90 degrees towards the viewer. The positions are also reflected from the left to the right of central star (similar to Fig.\,\ref{fg:ballistic-fit}) in order to aide comparison of the circular features. Assuming the nebula has a global spherical symmetry, this procedure allows us to compare the radial location of two key features: the narrow filaments and the infrared dusty nebula. We find strong evidence of a sharp, albeit jagged, boundary between the filaments and the bright, inner dust region. 
The filaments in Fig.\,\ref{fg:wise-overplot} are seen to cover the full radial extent of the outer nebula (dark purple region), where the gradient in infrared flux appears to be the strongest. The fact that the filaments do not reach the edge of the outer nebula along the x-axis (right ascension) is due to the footprint of the KCWI observations only extending out to the edge of the brightest [S\,{\sc ii}] filament, which reaches a maximum of RA=0h53m26s, and not beyond that (see Fig.\,\ref{fg:Sii-image}). 
\vspace{-10pt}

\section{Summary} \label{sec:summ}

We have here presented a detailed description of the velocity structure in the supernova remnant Pa~30 as revealed by the analysis of the Doppler-shifted [S\,{\sc ii}] 6717/6731\,\AA\ emission in the filaments of the nebula detected with KCWI-red. From our analysis, we can deduce that:
\begin{itemize}

    \item {\it The age of the nebula is consistent with Pa~30 being the remnant of SN 1181}. 
    This is consistent with previous studies \citep{ritter2021,fesen2023}. Our 3D analysis allows us to confirm that the velocities in the spectra are almost ballistic, and we can therefore derive the time of the supernova explosion to be $t_0=1152^{+77}_{-75}$\,yr.

    \item {\it The distribution of velocities in the ejecta is both narrow and highly symmetric}. As we can see from Fig.\,\ref{fg:velocity-histogram}, the velocity distribution is tightly peaked at about 1000\,km/s in both the blue- and red-shifted material. We find the maximum red- and blue-shifted velocities, defined as the 
    99\% confidence interval in the observed distribution, to be 1,440 and 1,380\,km\,s$^{-1}$ respectively, whilst the minima are 650 and 510\,km\,s$^{-1}$.

    \item {\it The ejecta show a strong asymmetry in flux along the line of sight, which may hint to an asymmetric explosion}. The red-shifted filaments are brighter and more numerous than the blue-shifted ones, even if we would expect the opposite to be true due to selection effects. We find that the total flux from red-shifted filaments is 40\% higher than the flux from blue-shifted ones. This is tantalizing evidence for asymmetry in the explosion; however, given the rather small area covered by our IFU observations, it may not be significant.
    Further IFU spectroscopic observations with wider coverage of the nebula will confirm if there exists a global asymmetry in the nebula ejecta, providing important constraints on dynamical models of the ejecta.
    
    \item {\it We confirm the presence of a large cavity in the remnant.} All of the filaments that appear close to the central star in the plane of the sky show very high line-of-sight velocities, implying that their line-of-sight separation from the star is large. The presence of a cavity had previously been inferred from long-slit spectroscopic observations\citep{fesen2023}. The KCWI observations presented here have allowed the detection of 
    a sharp inner edge to the filamentary emission in the ejecta, that coincides with the outer edge of the bright ring detected in the infrared. 
    Adopting the \textit{Gaia} distance of 2.3\,kpc, we measure the inner and outer radius of the filamentary shell to be $r_{\rm in}\approx0.6$\,pc and $r_{\rm out}\approx1.0$\,pc, respectively.
    The edge of this inner infrared ring has previously been interpreted to be indicative of the position of the reverse shock, as in the shocked region the dust grains should have been sublimated and destroyed \citep{ko2023,duffell2024}.
    \item {\it The velocities in the filaments are consistent with being ballistic}. We estimate the ballistic fraction in the ejecta velocities to be $k=0.97^{+0.09}_{-0.08}$. With $k$ very close to unity, there is no significant evidence that deceleration or acceleration has occurred in the ejecta.
    If the observed edge between the infrared cavity and the filaments indicates the location of the reverse shock, then we expect a large fraction of the ejecta to have been shocked already; the low deceleration observed would still be consistent with this scenario, and it would indicate a low density in the CSM. In the model by \citet{duffell2024}, for example, the deceleration induced by the reverse shock leads to a ballistic fraction of 0.93, consistent with our measurement.

\end{itemize}

\section*{Acknowledgements}
We thank Rob Fesen for providing the [S\,{\sc ii}] narrowband imaging and providing helpful comments on the paper. We also thank Eliot Quartert for helpful discussions.
TC was supported by NASA through the NASA Hubble Fellowship grant HST-HF2-51527.001-A awarded by the Space Telescope Science Institute, which is operated by the Association of Universities for Research in Astronomy, Inc., for NASA, under contract NAS5-26555. IC was also supported by NASA through grants from the Space Telescope Science Institute, under NASA contracts NASA.22K1813, NAS5-26555 and NAS5-03127.  
This research was supported in part by grant NSF PHY-1748958 to the Kavli Institute for Theoretical Physics (KITP).
OT was supported by FONDECYT grant 11241186.

This publication makes use of data products from the Wide-field Infrared Survey Explorer, which is a joint project of the University of California, Los Angeles, and the Jet Propulsion Laboratory/California Institute of Technology, funded by the National Aeronautics and Space Administration.

This research made use of Montage. It is funded by the National Science Foundation under Grant Number ACI-1440620, and was previously funded by the National Aeronautics and Space Administration's Earth Science Technology Office, Computation Technologies Project, under Cooperative Agreement Number NCC5-626 between NASA and the California Institute of Technology.

%

\vspace{5mm}
\facilities{Keck:II (KCWI)}


\software{\textsc{astropy} \citep{2013A&A...558A..33A,2018AJ....156..123A},  
\textsc{scipy} \citep{scipy2020}, \textsc{QFitsView}} \citep{ott2012-qfitsview}, \textsc{Montage} \citep{jacob_montage_2010} 




\appendix

\section{Details on data reduction and methodology}
\label{sec:app}
We reduced the raw KCWI data using the official \textsc{python} KCWI Data Reduction Pipeline\footnote{\url{https://github.com/Keck-DataReductionPipelines/KCWI_DRP}} (DRP) \citep{neill2023}. The DRP reads in afternoon calibration frames, science exposures, and standard star observations to produce flux-calibrated 3D data cubes of the science targets. The default parameters use ThAr arc spectra to construct a wavelength solution along with a master flat field derived from an average stack of internally illuminated flats.

In general, a raw science exposure is bias subtracted and gain corrected; cosmic rays are flagged and removed; the wavelength solution derived from calibration frames is applied; the frame is flat fielded; and a 1D sky model to the sky pixels in the spectra is subtracted from the data. At this point, the 2D data in detector coordinates is transformed into a 3D data cube, DAR corrected, and flux calibrated using the standard star observation. 

Although observations from both the blue and red channels undergo the same broad reduction steps, as mentioned above, cosmic ray hits are significantly more pernicious on the red detector. While \textsc{astroscrappy} \citep{astroscrappy,vanDokkum_CRR} is largely effective for the blue channel ``hot-pixel''-type cosmic rays, the red channel features a much more diverse array of cosmic ray morphologies that current automated routines are unequipped to handle. Instead, we opt to take several exposures and use their median to make cosmic ray masks for each constituent exposure. 

In practice, given three exposures of the same target under the same conditions, we construct a median exposure which effectively removes the cosmic rays and leaves just the science target. For each constituent exposure, we subtract off the median science frame we just constructed; the difference image then contains only cosmic rays plus some noise. We apply sigma-clipping to the difference image and convert the clipped pixels into a binary cosmic ray mask. We then run a customized version of the KCWI DRP\footnote{\url{https://github.com/prusinski/KCWI_DRP}} that reads in these cosmic ray masks, flags the affected regions, and excludes such pixels from the final science outputs. 

KCWI DRP processed cubes are stacked and mosaicked using \textsc{KCWIKit}\footnote{\url{https://github.com/yuguangchen1/KcwiKit}} \citep{prusinski2024}. The full implementation is described in \citep{chen2021}, but briefly, the individual science cubes are drizzled onto a common spatial grid with $0.3\arcsec\times 0.3\arcsec$ sampling using the \textsc{Montage} \citep{jacob_montage_2010} package. Each frame is also World Coordinate System (WCS) corrected using an external narrowband image.

To extract the Doppler information from the [S\,{\sc ii}] doublet in the mosaicked KCWI-red data, we create a mask to select only the brightest filaments (see Fig.\,\ref{fg:Sii-image}). The regions in the mask are determined by taking the mean flux average over the full data cube along the spectral dimension, between $\lambda=6800-6750\,\AA$, and subtracting the mean local continuum. This allows us to make a high-contrast image focused on all pixels with significant [S\,{\sc ii}] emission (lower panel of Fig.\,\ref{fg:Sii-image}). The wavelength range was chosen to include the minimally and maximally Doppler shifted doublets, with wavelength limits determined from visual inspection using the \textsc{QFitsView} software \citep{ott2012-qfitsview}. This high-contrast image was subsequently turned into an initial binary mask by selecting the brightest 30\% of pixels, after the exclusion of two bright and saturated stars (one being the central remnant, the other being located at RA=0h\,53m\,19s and Dec=67$^{\circ}$\,29'\,50''). This corresponds to a cut on mean fluxes less than $\langle F_{\rm [S\,{\sc ii}]} \rangle$\,$<$\,0.001\,$\times 10^{-8}$\,erg/s/cm$^{3}$/arscsec$^{2}$. This flux limit is about 4\% of the maximum flux in the high contrast image. After the masking of the two saturated stars, the spectral cube contained 25\,580 spatial pixels within the observed footprint. The flux cut on the high-contrast image yielded 8084 spatial pixels selected for analysis. We only analyze pixels centered on a 3$\times$3 pixel window which is fully contained within the selected regions. This was done in order to further reduce contamination from noise, and ensure a homogeneous analysis across the full cube. This reduced the total number of analyzed spatial pixels from 8084, to 5575. The number of pixels which received a single, double, triple, and quadruple doublet fit are 2681, 2469, 416, and 9, respectively. This yielded a total number of 8903 fitted [S\,{\sc ii}] doublets.



\bibliography{Pa30KCWI}{}
\bibliographystyle{aasjournal}



\end{document}